\documentclass[%
superscriptaddress,
 preprint,
 showpacs,preprintnumbers,
amsmath,amssymb,
aps,
 prb,
]{revtex4-1}
 
\usepackage{graphicx}% Include figure files
\usepackage{dcolumn}% Align table columns on decimal point
\usepackage{bm}% bold math
\usepackage{hyperref}% add hypertext capabilities
\hypersetup{
    colorlinks=true,    % 启用颜色链接
    linkcolor=blue,     % 设置内部链接（如 \ref）的颜色为蓝色
    citecolor=blue,     % 设置引用链接的颜色为蓝色
    filecolor=blue,     % 设置文件链接的颜色为蓝色
    urlcolor=blue       % 设置 URL 链接的颜色为蓝色
}
\begin{document}

%\preprint{APS/123-QED}

\title{Quasi-Bound States in the Continuum-Induced Second Harmonic Generation Enhancement in \texorpdfstring{High-$Q$}{High-Q} Triple Dielectric Nanoresonators}
% Force line breaks with \\

\author{Xu Tu}
\affiliation{Institute for Advanced Study, Nanchang University, Nanchang 330031, China}

\author{Meibao Qin}
\affiliation{School of Education, Nanchang Institute of Science and Technology, Nanchang 330108, China}

\author{Huifu Qiu}
\affiliation{Institute for Advanced Study, Nanchang University, Nanchang 330031, China}

\author{Feng Wu}
\affiliation{School of Optoelectronic Engineering, Guangdong Polytechnic Normal University, Guangzhou 510665, China}

\author{Tingting Liu}
\email{ttliu@ncu.edu.cn}
\affiliation{School of Information Engineering, Nanchang University, Nanchang 330031, China}
\affiliation{Institute for Advanced Study, Nanchang University, Nanchang 330031, China}

\author{Lujun Huang}
\email{ljhuang@phy.ecnu.edu.cn}
\affiliation{School of Physics and Electronic Science, East China Normal University, Shanghai 200241 China}

\author{Shuyuan Xiao}
\email{syxiao@ncu.edu.cn}
\affiliation{School of Information Engineering, Nanchang University, Nanchang 330031, China}
\affiliation{Institute for Advanced Study, Nanchang University, Nanchang 330031, China}

\begin{abstract}
	
High-$Q$ optical nanocavities are fundamental to modern optics and photonics, enabling enhanced light-matter interactions. Previous studies have demonstrated that high-$Q$ supercavity modes can be constructed within a single dielectric resonator by leveraging quasi-bound states in the continuum. However, their $Q$-factors are limited to few tens or hundreds when such a resonator is subwavelength scale. Here, we propose a general recipe for achieving high-$Q$ resonances with $Q>10,000$ in triple subwavelength dielectric resonators. This is realized through destructive interference between two resonant modes, optimized by structural tuning. Multipole analysis confirms that destructive interference across radiation channels suppresses loss, forming the ultrahigh-$Q$ states. These resonances can be efficiently excited by azimuthally polarized light due to improved mode overlap. As a key application, we demonstrate efficient second harmonic generation under this excitation, achieving a conversion efficiency of 6.6\% at an incident intensity of 0.78 GW/cm$^2$. Our results may find exciting applications in developing ultracompact photonic devices with superior performance. 

%\begin{description}
%\end{description}
\end{abstract}

\maketitle

%\tableofcontents

\section{\label{sec:level1}Introduction}

Nonlinear optics investigates the nonlinear phenomena arising from the interaction between light and matter, fundamentally governed by the nonlinear polarization response of a medium under high optical intensity \cite{Boyd2023}. In recent years, the rapid advancement of nanoresonators has opened up novel avenues for exploring nonlinear optical processes. Among various nanostructures, high-refractive-index nanoresontors have emerged as essential building blocks for manipulating light-matter interactions, owing to their ability to support multipolar Mie resonances with low optical losses \cite{smirnova2016multipolar,Bonacina2020,Grinblat2021,Sain2019}. When these nanoresonators are arranged into carefully designed architectures, their resonant modes can give rise to a variety of exotic optical states, such as Fano resonances \cite{Hopkins2013,Chong2014}, magnetic dipole (MD) resonances \cite{Feng2017,Deng2024,Smirnova2018}, toroidal dipole (TD) resonances \cite{Hasebe2020,MasoudianSaadabad2021,Sayanskiy2018}, nonradiating anapole states \cite{Miroshnichenko2015,Xu2018,Baryshnikova2019,Lu2022}, and bound states in the continuum (BICs) \cite{zhen2014topological,jin2019topologically,wu2019giant,hu2022global,kang2025janus}. These unique modes can significantly enhance local electromagnetic fields within ultra-small mode volumes, thereby providing an ideal platform for boosting the efficiency of nonlinear optical processes. Leveraging the strong localized field enhancement enabled by these resonant mechanisms, significant progress has been made across a variety of fields, including second-harmonic generation (SHG) \cite{Carletti2015,Makarov2017, Tu2024}, third-harmonic generation (THG) \cite{Shorokhov2016,Shcherbakov2014,MelikGaykazyan2017, Liu2025,sun2025high}, high-harmonic generation (HHG) \cite{Carletti2019,Zalogina2023}, frequency mixing\cite{Liu2018,liu2023high,Feng2023}, and nonlinear optical imaging \cite{Rodrigues2014,xu2019dynamic,RoscamAbbing2022}.

Among these resonant phenomena, BICs have attracted considerable attention due to their ultrahigh-quality ($Q$) factors. Recent studies have demonstrated that BICs can be realized even in a single nanoresonator \cite{Bogdanov2019,Rybin2017,Carletti2018,Koshelev2020,Volkovskaya2020,Odit2020,Huang2021}. These BICs are also known as Friedrich–Wintgen BICs, which arise from destructive interference between resonant modes that effectively cancels radiation leakage \cite{friedrich1985interfering,Hsu2016,huang2023resonant}. Ideally, a true BICs have infinite $Q$ factor. When radiative channels are weakly opened, BICs are transformed into quasi-BICs with finite yet extremely high-$Q$ factors, characterized by sharp resonance peaks and substantial local field enhancement \cite{koshelev2018asymmetric,Liu2019,Liu2021,dong2022schrodinger,wu2024momentum,liu2024efficient,zhou2024photonic,zhang2025strong}. This engineered control of radiation leakage has proven to be an effective strategy for enhancing the efficiency of nonlinear optical processes. Although quasi-BICs based on single nanoresonator supercavity modes have been extensively demonstrated, their relatively low-$Q$ factors and limited field enhancement restrict their applications. Some studies have reported quasi-BICs states with ultrahigh-$Q$ factors on the order of $10^4$ using single nanoresonator, but these typically rely on materials with extremely high refractive indices (such as $\varepsilon=80$) \cite{Rybin2017,Odit2020}, which hinders broader practical deployment. To overcome this, researchers have shifted toward exploring coupled resonant systems composed of multiple nanoresonators. By precisely controlling the position, spacing, and symmetry of multiple conventional dielectric nanoresonators (such as $\varepsilon=10$), novel quasi-BICs modes with high-$Q$ factors and extended electromagnetic field distributions include azimuthal polarized BICs formed via azimuthal symmetry breaking \cite{Deng2024}, radial BICs enabled by radial symmetry breaking \cite{Kuehner2022}, and chain-type BICs constructed in coaxial multi-nanoresonator arrays \cite{Nazarov2025,Sadrieva2019,bulgakov2017topological}. However, these designs typically require a large number of nanoresonators, ranging from dozens to theoretically infinite periodic arrays—which, despite enabling ultra-high-$Q$ factors, significantly complicate device fabrication and limit integration potential.

In this work, we propose a compact all-dielectric nanoresonators composed of only three subwavelength nanocuboids, engineered to support an ultrahigh-$Q$ quasi-BICs mode with $Q>10,000$. We show that ultrahigh-$Q$ resonances are enabled by constructing multiple avoided mode crossing via tuning four structural parameters. Multipolar decomposition suggests that destructive interference among different radiation channels leads to the formation of these ultrahigh quasi-BICs modes. We also evaluate the coupling efficiency between such high-$Q$ eigenmodes and various excitation wave (i.e. linear polarized (LP) wave and azimuthally polarized (AP) wave). Finally, we successfully demonstrate efficient second harmonic generations based on this high-$Q$ resonance by applying AP beam illumination. Thanks to the large overlapping between eigenmodes and AP beam, the conversion efficiency can be up to 6.6\% under AP beam excitation with a peak incident intensity of 0.78 GW/cm$^2$. We envision that such a compact platform can also be used to enhance light-matter interactions across both the linear and nonlinear regimes. 

\begin{figure}[htbp]
	\centering
	\includegraphics[width=0.5\textwidth]{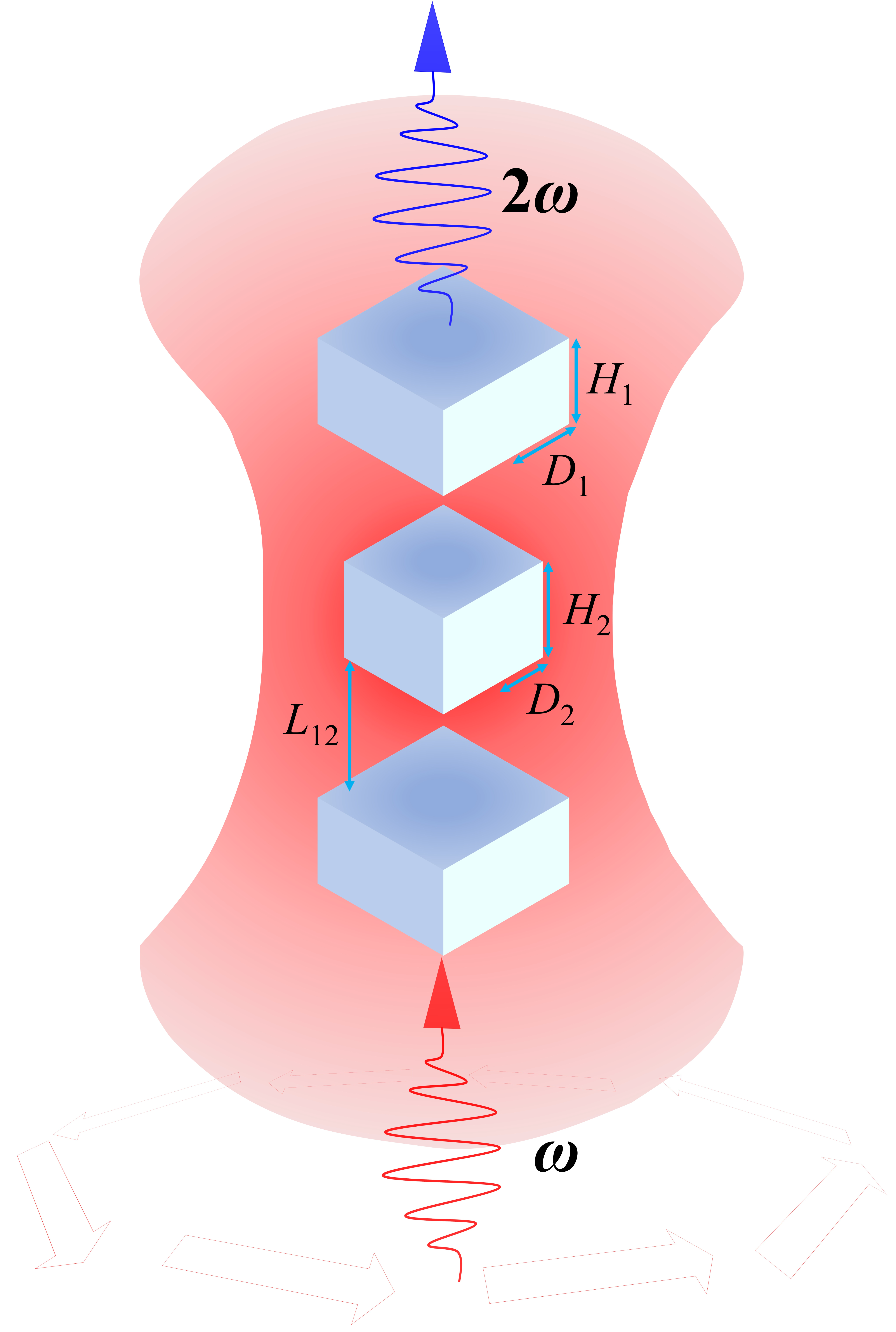}  % 图片文件名
	\caption{Schematic illustration of second harmonic generation in a triple nanocuboids AlGaAs system. Under illumination by an AP vector beam at frequency $\omega$, second harmonic signals at frequency 2$\omega$ are generated.}
	\label{fig1}
\end{figure}

\begin{figure*}[htbp]
	\centering
	\includegraphics[width=0.95\textwidth]{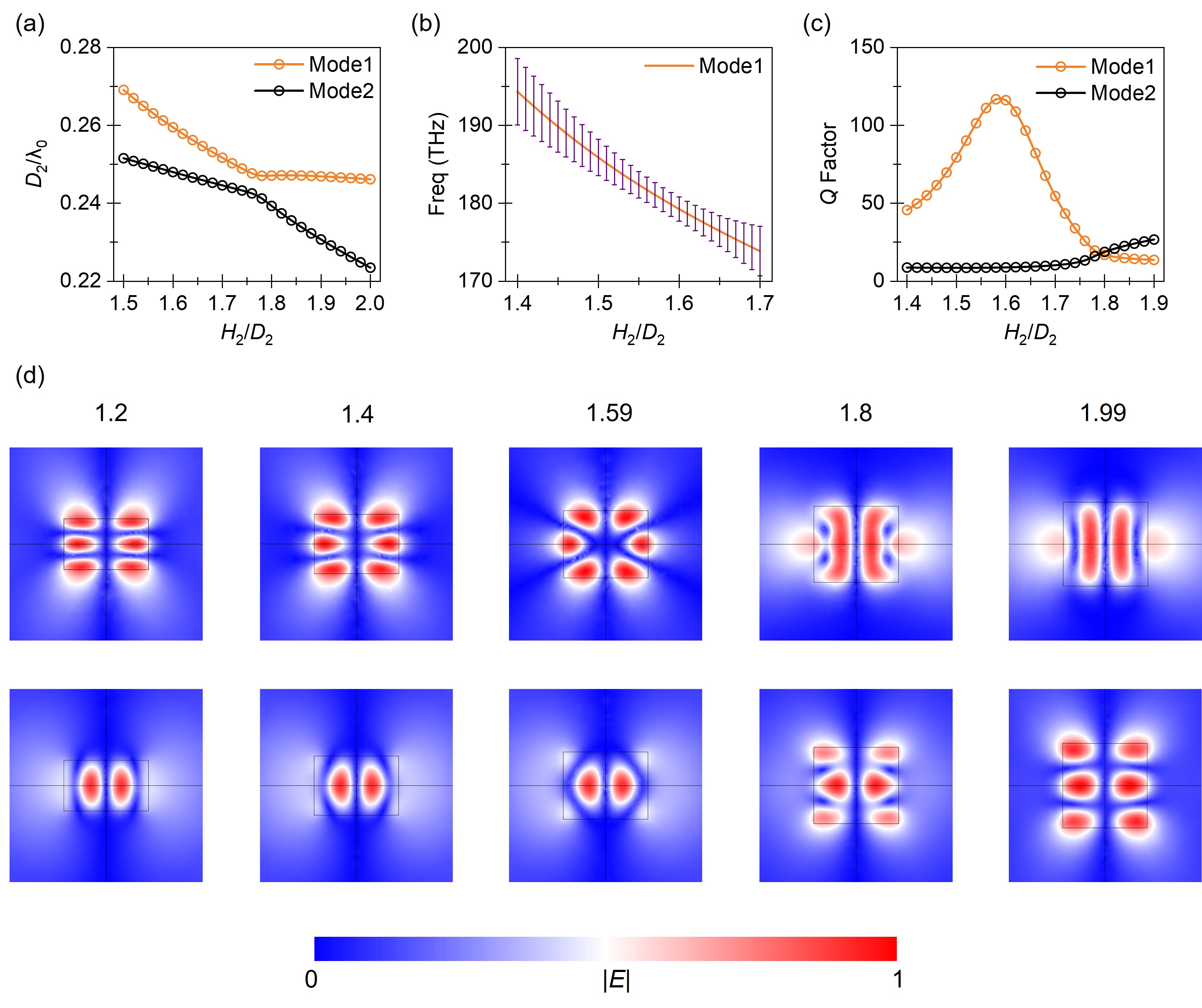}  % 图片文件名
	\caption{Eigenmode analysis of a single nanocuboid system. (a) Dispersion relation of a single nanocuboid system. Mode 1 corresponds to the high-$Q$ mode, and Mode 2 corresponds to the low-$Q$ mode. $\lambda_0$ denotes the resonant wavelength. (b) Variation of the resonant frequency and associated radiative loss of a single nanocuboid as a function of $H_2/D_2$. (c) Quality factors of the two modes. (d) Mode profiles of the two resonances in the $y$–$z$ plane. The numbers indicated in the figure correspond to the respective values of $H_2/D_2$.}
	\label{fig2}
\end{figure*}

\section{\label{sec:level2}Design and Modal Analysis of \texorpdfstring{High-$Q$}{High-Q} Dielectric Nanoresonators}

Here we focus on realizing quasi-BICs resonances with ultrahigh-$Q$ factor in a compact optical nanocavity. As illustrated in Fig. \hyperref[fig1]{1}, we consider a coaxially aligned system of three subwavelength AlGaAs nanocuboids arranged with mirror symmetry along the $z$-axis. The structure consists of one central nanocuboid flanked by two identical side nanocuboids. The central cuboid has a square base of side length $2D_2$ and height $H_2$, while the side cuboids each have a base of side length $2D_1$ and height $H_1$, with a center-to-center separation of $L_{12}$ from the central cuboid. Taking the base length of the central cuboid ($2D_2$) for normalization, the structure is characterized by four independently tunable dimensionless geometric parameters: the relative side length of the outer cuboids ($D_1/D_2$), their relative height ($H_1/D_2$), the relative height of the central cuboid ($H_2/D_2$), and the relative separation between cuboids ($L_{12}/D_2$). By optimizing these parameters, we successfully construct multiple avoided crossings in the mode dispersion diagram—that is, the relationship between the normalized mode frequency and the geometric parameters—which in turn significantly enhances the $Q$ factor of the resulting resonant modes.

To better understand the origin of these high-$Q$ resonances, we first analyze the modal behavior of a single nanocuboid in isolation. A prominent feature of the normalized eigenfrequency ($D_{2}/\lambda_{0}$) shown in Fig. \hyperref[fig2]{2}(a) is the appearance of an avoided crossing when varying the aspect ratio $H_{2}/D_{2}$, indicating strong coupling between two resonant modes. This modal hybridization leads to destructive interference that are central to achieving high-$Q$ resonances in open photonic systems\cite{Bogdanov2019,Rybin2017,Carletti2018,Koshelev2020,Volkovskaya2020,Odit2020,Huang2021}. As shown in Figs. \hyperref[fig2]{2}(b) and 2(c), the radiative losses are significantly suppressed, and the $Q$ factor reaches a maximum when $H_{2}/D_{2}=1.59$. The mode corresponding to this peak is the so-called supercavity mode, also referred to as a quasi-BICs mode, with a calculated $Q$ factor of 117. Fig. \hyperref[fig2]{2}(d) further depicts the evolution of the mode profiles as a function of aspect ratio, highlighting the mode profile interchange across the avoided crossing. These insights provide the groundwork for understanding more complex multi-nanocuboids interactions, which we explore through a stepwise optimization framework in the following analysis.

\begin{figure}[htbp]
	\centering
	\includegraphics[width=0.75\textwidth]{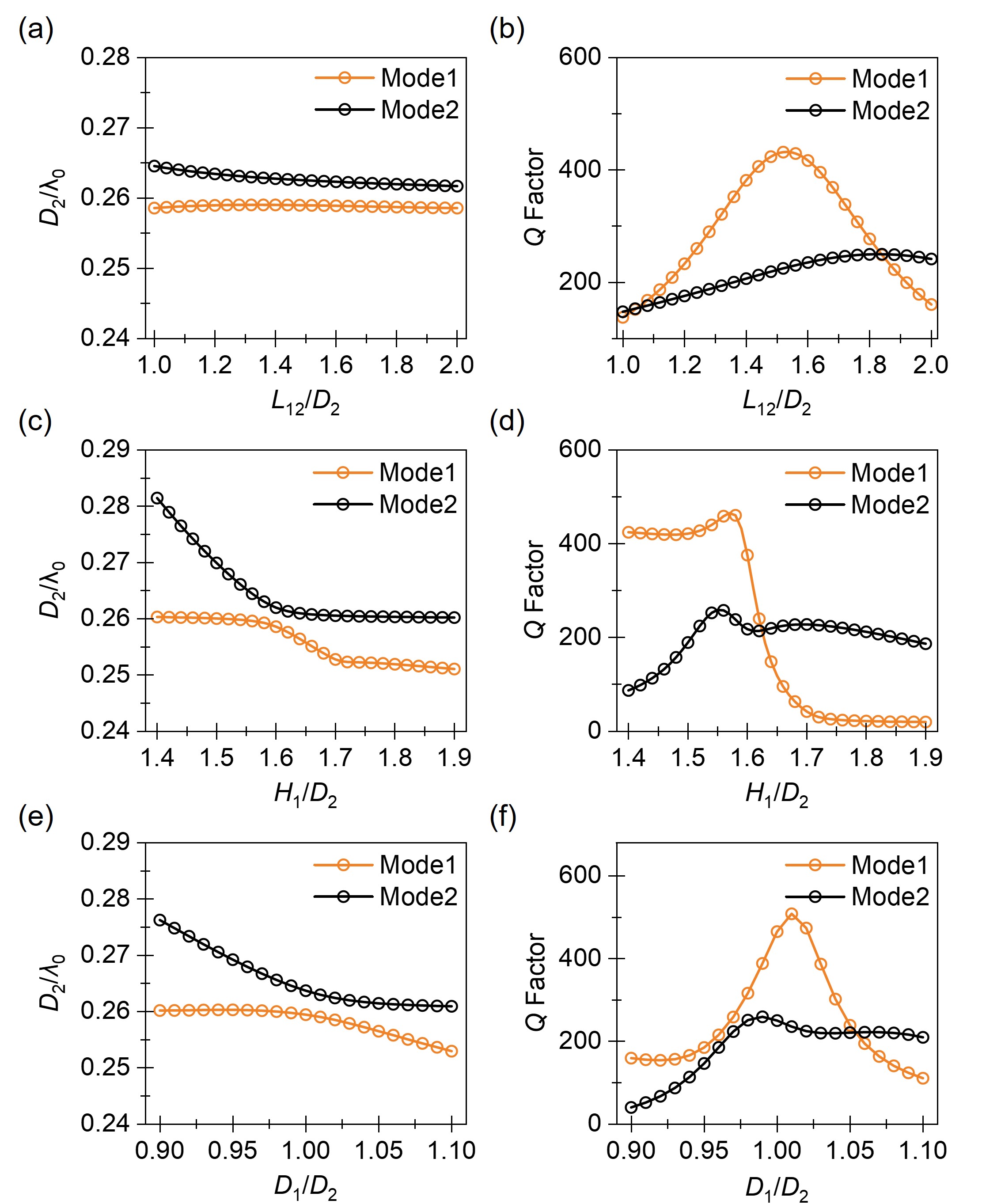}  % 图片文件名
	\caption{Evolution analysis of eigenmodes during parameter optimization. (a), (c), and (e) show the normalized frequencies ( $D_{2}/\lambda_{0}$) of two eigenmodes as functions of varying geometric parameters: (a) inter-cuboid distance $L_{12}/D_{2}$, (c) height of the outer nanocuboids $H_{1}/D_{2}$, and (e) base width of the outer nanocuboids $D_{1}/D_{2}$. (b), (d), and (f) display the corresponding quality factor curves for the two modes under each parameter variation, illustrating their radiative loss characteristics.}
	\label{fig3}
\end{figure}

Building on the eigenmode analysis of the single nanocuboid, we proceed to design and optimize a triple nanocuboids system. We describe this systematic parameter optimization process aimed at constructing high-$Q$ resonant modes. The core objective of the optimization is to further suppress radiative losses by engineering destructive interference among multipolar radiation channels through geometric tuning within a four-dimensional parameter space. The procedure is carried out in the following steps:

Step 1: We begin by optimizing the aspect ratio $H_{2}/D_{2}$ of a single nanocuboid to obtain its optimal resonant state. As shown in Fig. \hyperref[fig2]{2}, the $Q$ factor reaches a maximum value of $Q= 117$ when $H_{2}/D_{2} = 1.59$.

Step 2: Keeping the inner cuboid parameter fixed at $H_{2}/D_{2} = 1.59$, we introduce two outer nanocuboids with the same aspect ratio, i.e., $H_{1}/D_{1} = 1.59$, and set $D_{1}/D_{2} = 1$. The remaining optimization parameter in this step is the inter-nanocuboid distance $L_{12}$. As shown in Fig. \hyperref[fig3]{3}(b), when $L_{12}/D_{2} = 1.53$, the system reaches a new optimal resonance state with an enhanced quality factor of $Q = 432$, representing a 3.7-fold improvement compared to the initial configuration.

Step 3: With $H_{2}/D_{2}$ and $L_{12}$ held constant, we further optimize the height $H_{1}$ of the outer nanocuboids. As indicated in Fig. \hyperref[fig3]{3}(d), the $Q$ factor increases to $Q = 465$ when $H_{1}/D_{1} = 1.57$.

Step 4: Finally, we optimize the base width $D_{1}$ of the outer nanocuboids. As shown in Fig. \hyperref[fig3]{3}(f), the calculations reveal that when $D_{1}/D_{2} = 1.01$, the system reaches its overall optimal resonant state, with the $Q$ factor further enhanced to $Q = 508$.

This four steps process constitutes a full optimization cycle for achieving high-$Q$ resonant states through precise geometric engineering. Upon completion of one cycle, we re-adjust the aspect ratio $H_{2}/D_{2}$ of the inner nanocuboid and initiate a new iteration. After several such iterative optimization cycles, we arrive an optimal resonant state characterized by a substantially enhanced $Q$ factor, which exceeds that of the single nanocuboid counterpart by more than two orders of magnitude.

\begin{figure*}[htbp]
	\centering
	\includegraphics[width=0.95\textwidth]{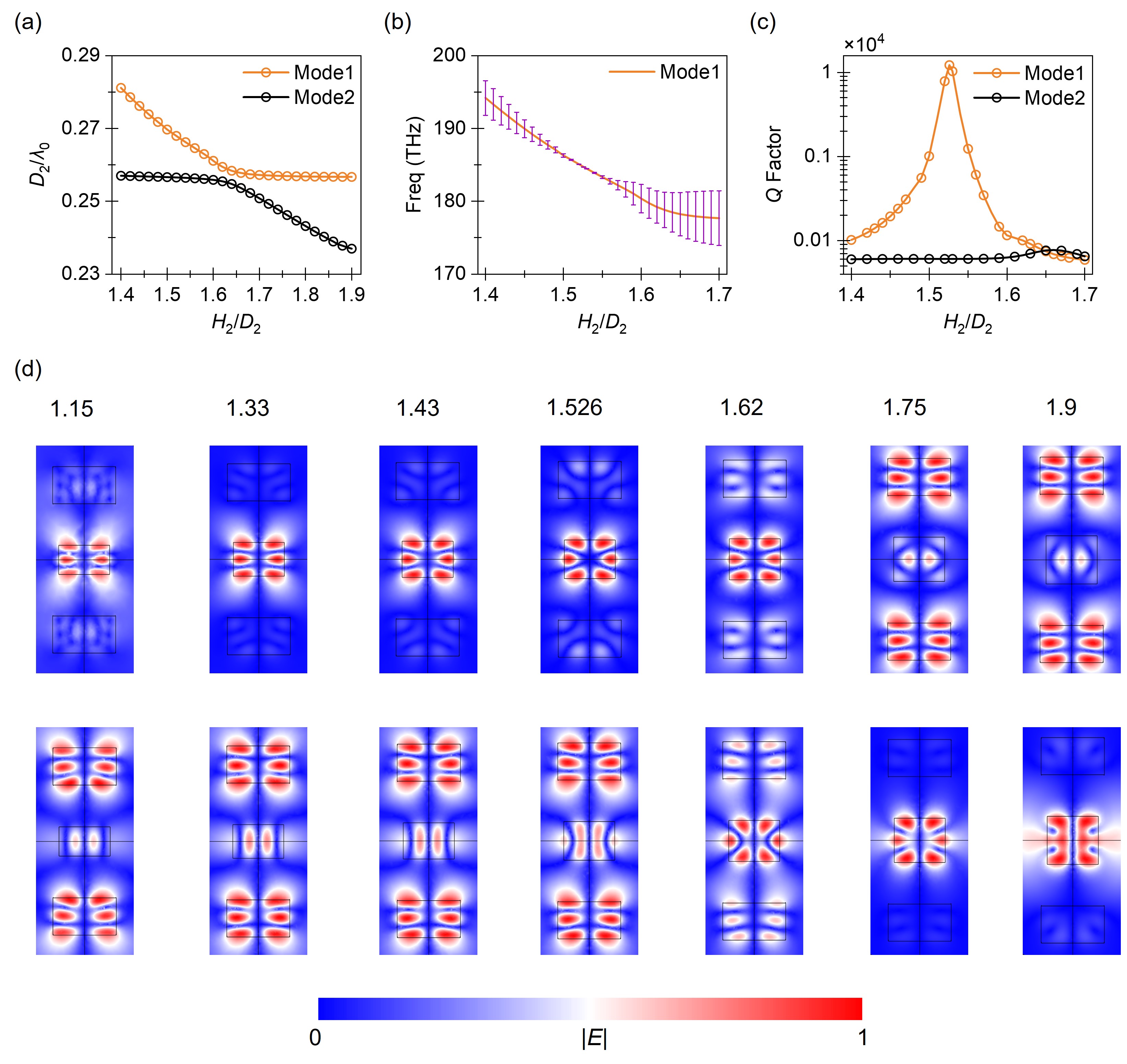}  % 图片文件名
	\caption{Eigenmode analysis of the triple nanocuboid system. (a) Dispersion relation of the triple nanocuboid system. Mode 1 corresponds to the high-$Q$ mode, and Mode 2 corresponds to the low-$Q$ mode. $\lambda_{0}$ denotes the resonant wavelength. (b) Variation of resonant frequency and corresponding radiative loss with respect to $H_{2}/D_{2}$ in the triple nanocuboids system. (c) $Q$ factors of the two modes. (d) Mode profiles of the two resonances in the $y$–$z$ plane. The numbers indicated in the figure correspond to the respective values of $H_{2}/D_{2}$.}
	\label{fig4}
\end{figure*}

After obtaining the optimized structural parameters ($H_{1}/D_{2} = 1.461$, $L_{12}/D_{2} = 1.636$, $D_{1}/D_{2} = 1.234$), we further examine the physical characteristics of the supercavity mode supported by the coaxial AlGaAs triple nanocuboids structure. Specifically, by keeping the parameters of the two outer nanocuboids fixed, we systematically study the influence of the central cuboid's aspect ratio ($H_{2}/D_{2}$) on the resonant frequency and $Q$ factor. As shown in Fig. \hyperref[fig4]{4}(a), the dispersion curve again exhibits a clear avoided crossing behavior. Fig. \hyperref[fig4]{4}(b) illustrates the variation of the resonance frequency and corresponding radiative loss of the high frequency mode with respect to $H_{2}/D_{2}$. It can be observed that when $H_{2}/D_{2} = 1.526$, the radiative loss is significantly suppressed, and the $Q$ factor reaches its maximum ($Q=12,115$). As $H_{2}/D_{2}$ deviates from this value, radiative leakage gradually increases. Similarly, Fig. \hyperref[fig4]{4}(d) presents the evolution of the eigenmodes, clearly visualizing the continuous transition between hybridized modes across the avoided crossing.

\section{\label{sec:level3}Multipolar Interference and Coupling Coefficients}
This section aims to elucidate the formation mechanisms and excitation of high-$Q$ resonant modes, as well as presents a theoretical analysis from the perspectives of far-field multipolar interference and excitation-coupling efficiency. On one hand, employing a multipolar decomposition in spherical coordinates, we calculate the multipole contributions and their relative phase relationships in the far-field radiation. Leveraging the principle of destructive multipolar interference, we reveal the crucial role of radiation suppression in enabling the formation of high-$Q$ resonances. On the other hand, by defining the modal field overlap integral, we quantitatively evaluate the coupling efficiency between incident sources with different polarization states and the eigenmodes. It is demonstrated that AP light, owing to its strong overlap with the transverse electric field components of the mode, can efficiently excite quasi-BICs modes.

\subsection{\label{sec:level3.1}Multipolar Interference and Cancellation}

In Section~\ref{sec:level2}, the design of high-$Q$ triple nanocuboids system is achieved through a four-parameter optimization scheme. Although this optimization process successfully find the $Q$ factor, the underlying physical mechanisms have not yet been thoroughly elucidated. To explain it, we exploit the multipolar decomposition in spherical coordinates to solve the far field electromagnetic radiation for different components of the nanocuboids separately \cite{Grahn2012,pichugin2019,Pichugin2021,Bulgakov2021,Pichugin2023}. The calculated results reveal that the high-$Q$ quasi-BICs mode arises due to destructive interference in the multipolar radiation from the different subsystems of the nanoresonators. Specifically, by leveraging the orthogonality of vector spherical harmonics $\vec{X}_{lm}$ and scalar spherical harmonics $\vec{Y}_{lm}$, we can calculate the electric $a_\text{E}(l,m)$ and magnetic multipole coefficients $a_\text{M}(l,m)$ by performing a surface integral over the entire spherical surface surrounding the scatterer (see Supplementary Material for details). These coefficients represent the amplitude and phase of the electric and magnetic multipolar radiation. The relative radiation intensity of each multipolar component, $P_{lm}$, can be obtained by normalizing the squared amplitude of the corresponding multipole coefficient. Specifically, it is defined as the ratio of the squared modulus of the multipole coefficient to the total radiated power, representing the fractional contribution of each multipolar channel

\begin{equation}
P_{{}_{ lm}} = \sqrt\frac{\left|a_{{}_{\rm E}} \left(l,m\right)\right|^{2}+ \left|a_{{}_{\rm M}} \left(l,m\right)\right|^{2}}{\sum\limits_{l=0}^{\infty} \sum\limits_{m=-l}^{l}\left(\left|a_{{}_{\rm E}} \left(l,m\right)\right|^{2}+ \left|a_{{}_{\rm M}} \left(l,m\right)\right|^{2}\right)}.
\end{equation}
In the considered coaxial nanocuboids system, due to the mirror symmetry about the $z$-axis, nonzero multipolar radiation components only arise when the orbital angular momentum quantum number $l$ is odd \cite{bulgakov2021exceptional}. Furthermore, since the electric field mode approximately exhibits axial symmetry along the $z$-axis, the system is invariant under rotations around the $z$-axis, which fixes the magnetic quantum number to $m = 0 $\cite{Pichugin2021}. 

\begin{figure}[htbp]
	\centering
	\includegraphics[width=0.75\textwidth]{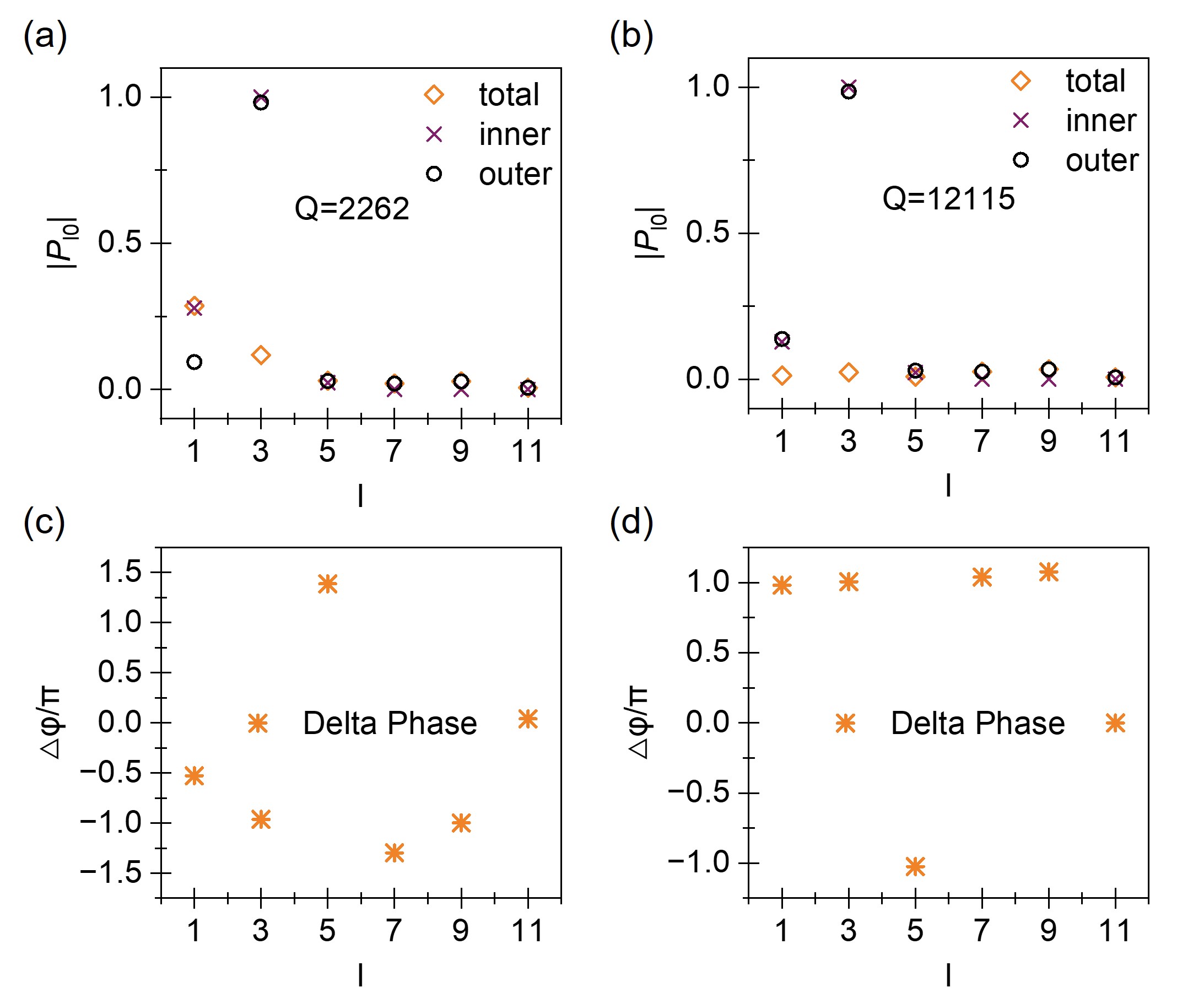}  % 图片文件名
	\caption{Comparison of multipolar decomposition results, showing the radiation contributions and phases of multipolar channels for modes with different $Q$ factors. Panels (a) and (b) plot the multipolar contributions of the total system (“total”), the middle nanocuboid (“inner”), and the outer nanocuboid pair (“outer”). Panels (c) and (d) correspond to the radiation phase difference $\Delta \phi/\pi$, representing the phase difference between the inner and outer subsystems for each multipolar channel. In this case, the reference side length $D_2$ is set to 434 nm. The relative side length of the outer cuboids is $D_1 / D_2 = 1.234$, the relative height of the outer cuboids is $H_1 / D_2 = 1.461$, and the relative spacing between the cuboids is $L_{12} / D_2 = 1.636$. For panels (a) and (c), the relative height of the middle cuboid is $H_2 / D_2 = 1.508$, while for panels (b) and (d), it is $H_2 / D_2 = 1.526$.}
	\label{fig5}
\end{figure}

To analyze the roles of different subsystems in multipolar interference, we divide the triple nanopcuboids into two subsystems: the middle nanocuboid is defined as the “inner” subsystem, while the top and bottom nanocuboids combined form the “outer” subsystem. Together, they constitute the total system, denoted as “total”. We calculate the relative multipole amplitudes and corresponding radiation phase differences for the triple systems (inner, outer, and total). The normalized multipolar relative radiated intensity is shown in Fig. \hyperref[fig5]{5}, where the relative contributions of the inner subsystem are marked by “×”, the outer subsystem by “$\circ$”, and the total system by “$\diamond$”. In contrast to the case in Figs. \hyperref[fig5]{5}(a) and {5}(c) with a $Q$ factor of 2262, where significant destructive interference is observed only in the $l=3$ channel, the high-$Q$ modes in Figs. \hyperref[fig5]{5}(b) and {5}(d) ($Q = 12,115$) demonstrate not only nearly complete destructive interference in this channel, but also significant suppression of radiation from higher-order multipolar channels (e.g. $l=5,7,9$). Multiple multipolar channels across different orders exhibit near perfect out-of-phase superposition, resulting in a substantial reduction of the total radiated power. This mechanism of multi-order destructive interference effectively reduces the system’s radiative losses and constitutes the fundamental physical origin of the formation of high-$Q$ modes. Notably, this interference behavior characterized by a strict phase control near $\pi$ across multiple multipolar channels is achieved through the synergistic optimization of structural parameters at different length scales within the system. This ensures that electromagnetic modes of different orders destructively interfere in the far field radiation, thereby maximizing the suppression of multipolar radiation and enabling the formation of quasi-BICs with extremely low radiation leakage.

\subsection{\label{sec:level3.2}Analysis of Coupling Coefficients}
After gaining the comprehensive understanding of the formation mechanism governing this quasi-BICs mode, it becomes essential to explore strategies for its efficient excitation, as the resultant optical response is critically dependent on the coupling efficiency between the incident light and the quasi-BICs state. Within high-$Q$ dielectric nanoresonators, the electric field distribution of the quasi-BICs mode manifests a characteristic ring-like pattern confined to the $x$-$y$ plane, exhibiting azimuthal polarization (i.e., along the angular direction). Vector beams—an archetype of structured light—have garnered significant attention in nanophotonics due to their unique capability for spatial polarization control \cite{Youngworth2000,wang2020generating,Buono2022}, with AP beams representing a specific category. The characteristic azimuthal polarization renders AP beams particularly suitable for the selective excitation of multipolar resonances. Previous research has demonstrated their efficacy in enhancing local electromagnetic fields and facilitating advances in nonlinear optics \cite{Carletti2015,Makarov2017,Shcherbakov2014,MelikGaykazyan2017,Carletti2019,Carletti2018,Koshelev2020,Volkovskaya2020}. To achieve optimal spatial and polarization matching with this profile, we utilize an AP beam as the excitation source. Our simulation results confirm that AP illumination effectively excites the quasi-BICs mode, yielding a significantly enhanced coupling efficiency compared to conventional excitation approaches. This observed enhancement motivates a quantitative assessment of the coupling efficiency. 

To quantitatively describe the coupling efficiency between different excitation sources and the quasi-BICs mode, we define a coupling coefficient that characterizes the field overlap between the pump beam and the resonant mode at the resonance wavelength as follows \cite{Carletti2018}
\begin{equation}
\eta_{{}_{i}}= \frac{\left|\int_{S}\vec{E}_{\rm inc}^{\ast} \cdot \vec{E} _{\rm mode}dS\right|^{2}}{\left(\int_{S} \left|\vec{E}_{\rm inc}\right|^{2}dS \right) \cdot \left(\int_{S} \left|\vec{E}_{\rm mode}\right|^{2}dS \right)}.
\end{equation}
Here, $\vec{E}_{\text{inc}}$ denotes the electric field of the incident pump light (including LP and AP sources), and $\vec{E}_{\text{mode}}$ represents the electric field distribution of the resonant mode. The integration domain $S$ corresponds to the top surface of the nanocuboids. The denominator serves as a normalization factor to ensure that the resulting coupling coefficient is dimensionless and comparable across different excitation conditions. Through numerical calculations, we obtain a coupling coefficient of $\eta_{\text{LP}} = 2.95 \times 10^{-10}$ for LP light, whereas the coefficient for AP light is $\eta_{\text{AP}} = 0.38$. These results indicate that the coupling efficiency between AP light and the quasi-BICs mode is significantly higher than that of LP light. This phenomenon can be reasonably explained from a symmetry perspective: the quasi-BICs mode exhibits even $C_2$ symmetry in the $x$–$y$ plane, while LP light possesses odd $C_2$ symmetry. As a result, their modal overlap is nearly zero, rendering LP light ineffective in exciting the quasi-BICs resonance. In contrast, AP light not only shares a high degree of spatial overlap with the mode field distribution, but its polarization direction also closely matches the azimuthal electric field orientation of the quasi-BICs mode. This leads to a significantly enhanced overlap integral and thus a much stronger excitation efficiency. Therefore, from both the viewpoint of symmetry and polarization matching between the excitation source and the mode field, and from the quantitative comparison of coupling coefficients, AP light demonstrates superior performance as an excitation source for quasi-BICs resonances. Accordingly, we employ AP light as the background excitation field in our simulations, which enables a significantly enhanced SHG efficiency.
 
\section{\label{sec:level4}Optical Response of the Optimized High-\texorpdfstring{$Q$}{Q} Nanoresonators}

Next, we proceed to analyze the optical responses of the designed high-$Q$ triple nanoresonators, including both the linear and nonlinear regimes. The linear response analysis focuses on the manifestation of the resonant characteristics, including resonance enhancement in the scattering spectrum, the physical origin of multipolar contributions, and the symmetry properties of the resonant modes. The nonlinear response, on the other hand, explores the capability of the resonant state to enhance nonlinear processes, particularly its performance in SHG. It is worth noting that although our previous eigenmode optimization procedure successfully yielded a structure with an ultrahigh-$Q$ factor at a specific parameter set, for the following analyses linear and nonlinear response analyses concerning realistic excitation scenarios, we deliberately tuned the strucural parameters (see Supplementary Material for details) to achieve moderate $Q$ factors in the range of approximately 2,000. This strategic shift is driven by the fundamental need to optimize coupling efficiency under practical plused excitation. As noted in studies of nonlinear frequency conversion in resonant systerms \cite{anthur2020continuous,liu2021giant}, the substantial spectral width ($\sim$ 1 nm typical) of a pulsed pump laser presents a critical mismatch with the narrow linewidth of ultrahigh-$Q$ resonances. This mismatch leads to a significant portion of the pump energy residing outside the resonant mode bandwidth and thus not contributing to the target nonlinear process, such as SHG, potentially causing experimentally observed efficiencies to fall considerably below theoretical predictions. To maximize the overlap between the pump bandwidth and the resonant linewidth and thereby ensure maximal energy transfer from the pulsed pump source to the mode for efficient nonlinear generation, we specifically targeted this optimal $Q$ factor range ($Q$ $\sim$ 2,000, commensurate with the $\sim$ 1 nm laser bandwidth). This approach focuses the analysis on practically relevant excitation conditions while enhancing the direct applicability of the resonant designs. Through this comprehensive analysis, we not only reveal that the high-$Q$ structure supports quasi-BICs-type resonances dominated by magnetic multipoles, but also quantify its nonlinear radiation efficiency under AP beam excitation. This provides both theoretical insight and design guidance for the application of high-$Q$ resonant structures in enhanced nonlinear optical devices.

\subsection{\label{sec:level4.1}Linear Optical Response}

\begin{figure}[htbp]
	\centering
	\includegraphics[width=0.75\textwidth]{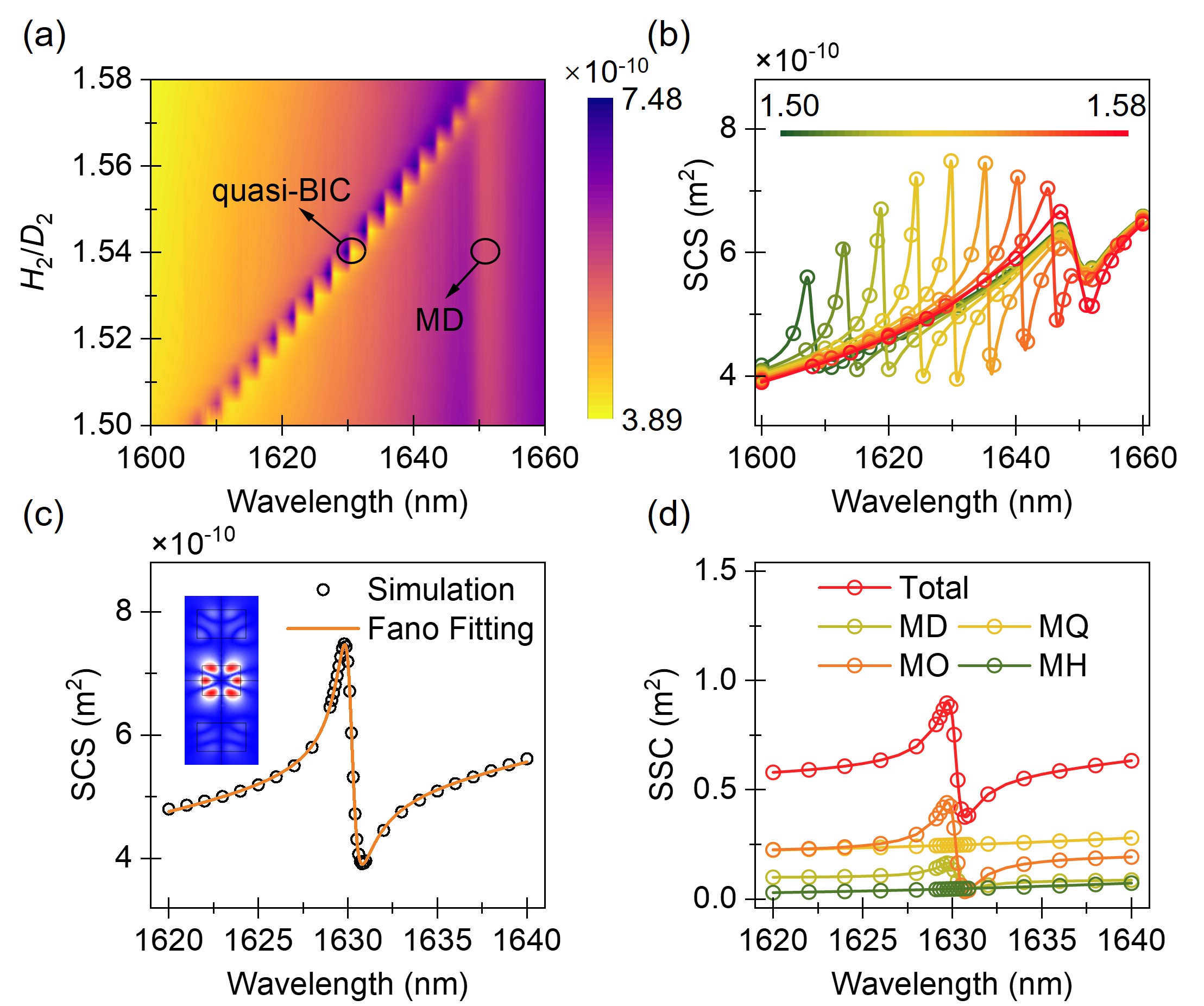}  % 图片文件名
	\caption{Linear optical response of the triple nanocuboid system. (a) Variation of the scattering cross section as the parameter $H_2/D_2$ increases. (b) Comparison of scattering cross sections for different structural parameters $H_2/D_2$. (c) Fitting of the Fano resonance formula. The scatter plot represents the simulated data, and the solid curve shows the fitted result using the Fano resonance model with selected parameters. The inset shows the mode profile of the structure in the $y$-$z$ plane under AP beam excitation. (d) Contributions of various multipole orders to the scattering cross section calculated via spherical multipole decomposition, including MD, MQ, MO, and magnetic hexadecapole (MH); the cross section values are normalized.}
	\label{fig6}
\end{figure}

We start from the linear optical response characteristics of the AlGaAs nanocuboids, with a particular focus on exciting the quasi-BICs mode obtained from the eigenmode solver. We perform three-dimensional full-wave electromagnetic simulations using the frequency domain solver in COMSOL Multiphysics to analyze the optical properties of the structure. In the simulations, the entire model is placed in an air ($n=1$) environment, and the background field is illuminated by an AP beam (see Supplementary Material for details). The refractive index of the scattering nanocuboids is taken from experimentally reported data in the literature \cite{Gehrsitz2000}. The scattering cross-section is quantitatively evaluated using the following equation

\begin{equation}\sigma_{\mathrm{sca}}=\frac{1}{I_0}\oint_S\left(\vec{E}_{\mathrm{sca}}\times\vec{H}_{\mathrm{sca}}^*\right)\cdot\vec{n}dS,\end{equation}
Here, $\vec{n}$ denotes the outward-pointing normal vector from the nanocuboid surface, and $I_0$ represents the incident light intensity. Fig. \hyperref[fig6]{6}(a) shows the wavelength-dependent scattering cross-section $\sigma_{\text{sca}}$ as a function of the $H_2/D_2$ parameter. The results indicate that near the resonance wavelength corresponding to each $H_2/D_2$ value, the scattering cross-section varies drastically with wavelength, and the resonance peak position is highly dependent on geometric parameters. To quantitatively analyze this parameter sensitivity, Fig. \hyperref[fig6]{6}(b) compares the scattering spectra under different $H_2/D_2$ ratios. It can be observed that when $H_2/D_2 = 1.54$, the system reaches its optimal resonant condition, characterized by the highest peak scattering cross-section ($\sigma_\text{{sca}}^\text{{max}} = 7.48\times10^{-10}$ m²) and the smallest full width at half maximum (FWHM = 0.966 nm). As the parameter value deviates from this optimal point, the resonance strength gradually weakens and the linewidth broadens.

Notably, a persistent MD resonance mode is observed near the 1650 nm, which is independent of the central nanocuboid’s geometry. Its near-field distribution is primarily localized within the two side nanocuboids. This phenomenon arises from the geometric resonance effect induced by the specific aspect ratio of the side nanocuboids. This mode can also enhance the local field and can be effectively excited by the AP beam. However, compared to the quasi-BICs mode ($Q = 1,958$, field enhancement factor = 30), the MD resonance exhibits significantly lower field enhancement and a lower quality factor ($Q = 309$, field enhancement factor = 8). To analyze the spectral line shape of the resonance, we perform a Fano resonance fitting for a representative set of structural parameters, as shown in Fig. \hyperref[fig6]{6}(c). Around the resonance wavelength, a standard Fano formula fitting is applied to the simulated scattering cross-section data using the Levenberg–Marquardt nonlinear least-squares algorithm \cite{Bogdanov2019,Limonov2017}

\begin{equation}\sigma(\omega)=\frac{c^2}{\omega^2\left|E_{\mathrm{inc}}\right|^2}\left[\frac{A}{1+q^2}\frac{\left(q+\Omega\right)^2}{1+\Omega^2}+I_{\mathrm{bg}}\left(\omega\right)\right],\end{equation}
Here, $A$ represents the peak amplitude, $q$ is the Fano asymmetry parameter, $\Omega=(\omega-\omega_0)/\gamma$, $\gamma$ is the leakage rate \cite{xu2019dynamic,koshelev2019nonlinear,wu2019giant}, and $I_{bg}$ is the background term of the scattering cross-section. The fitting results reveal a resonance wavelength $\lambda_0 = 1630.2$ nm and linewidth $\Gamma = 0.966$ nm, and a corresponding quality factor $Q = \lambda_0/\Gamma\approx1,688$, confirming the presence of strong Fano interference in the system. At the resonance wavelength, we plotted the mode profile of the structure under AP beam excitation and found it to be highly consistent with the mode obtained from the eigenmode solver, with only slight up-down asymmetry. This demonstrates that the quasi-BICs mode can be efficiently excited by the AP beam. Furthermore, a quantitative analysis of the background scattering field is conducted via spherical harmonic multipolar decomposition. The results show that the dominant contribution to the scattering field arises from higher-order magnetic multipole modes, with the MO component accounting for up to 49\% of the total. This finding is consistent with the spatial field distribution characteristics of the quasi-BICs mode.

\begin{figure*}[htbp]
	\centering
	\includegraphics[width=0.95\textwidth]{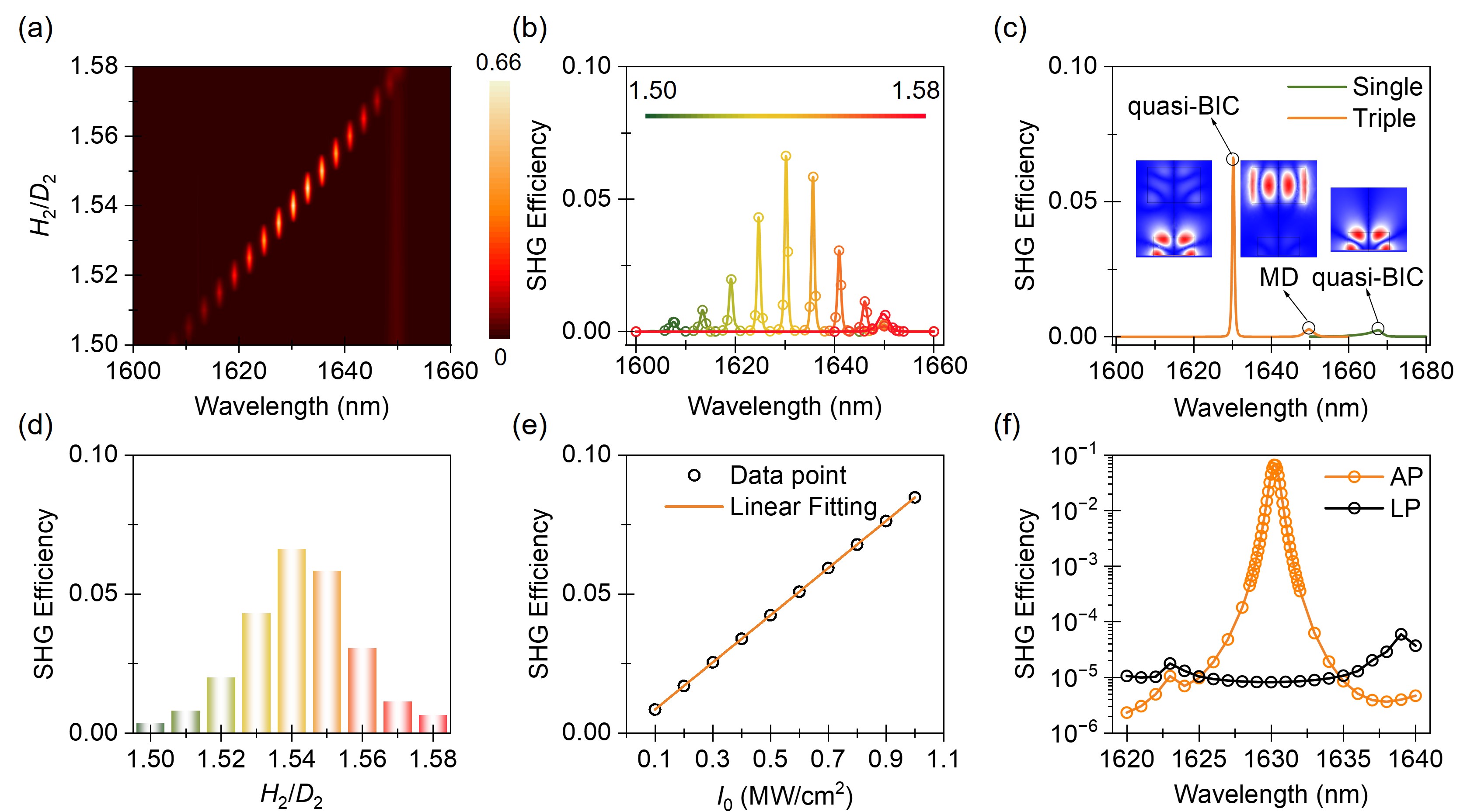}  % 图片文件名
	\caption{Nonlinear optical response of the triple nanocuboid system. (a) Variation of SHG efficiency as the parameter $H_2/D_2$ increases. (b) Comparison of SHG efficiency for different structural parameters $H_2/D_2$. (c) SHG efficiency at a specific value of $H_2/D_2$. (d) Dependence of SHG efficiency on the incident pump power. (e) Wavelength-dependent SHG efficiency comparison between the triple nanocuboids and the single nanocuboid. The inset shows the eigenmode electric field distribution in the $x$–$y$ plane at the resonance wavelength. (f) SHG efficiency as a function of wavelength under AP and LP illumination when $H_2/D_2 = 1.54$.}
	\label{fig7}
\end{figure*}

\subsection{\label{sec:level4.2}Nonlinear Optical Response}

We then turn our attention to the quasi-BICs-induced nonlinear response, applying the optimized structure to enhance SHG. The coupled-wave equations at the fundamental and second-harmonic wavelengths can be written as \cite{saleh2019fundamentals}

\begin{equation}
\vec{\nabla}^{2}\vec{E}-\frac{1}{c^{{}^{2}}}\frac{\partial^{2}\vec{E}}{ \partial t^{{}^{2}}}=\mu_{{}_{0}}\frac{\partial^{2}\vec{P}_{{}_{\rm nl}}}{ \partial t^{{}^{2}}},
\end{equation}
where $c = c_0 / n$, $n = 1 + \chi$, and $c_0 = 1/\sqrt{\epsilon_0 \mu_0}$ is the speed of light in vacuum. Here, $\chi$ is the linear electric susceptibility, $\vec{E}$ is the electric field generated by the nonlinear polarization source, and $\vec{P}_{\text{nl}}$ is the nonlinear polarization. In the SHG process, the nonlinear polarization $\vec{P}_{\text{nl}}$ is typically represented as the second-harmonic nonlinear polarization $\vec{P}_{\text{nl}}^{(2\omega)}$. For AlGaAs material, it is given by (see Supplementary Material)
\begin{equation}
\begin{pmatrix}P_{nlx}^{ \left( 2 \omega \right)}\\ P_{nly}^{ \left( 2 \omega \right)}\\ P_{nlz}^{ \left( 2 \omega \right)} \end{pmatrix}=2 \epsilon_{0} \chi^{ \left( 2 \right)} \begin{pmatrix}E_{y}^{ \left( \omega \right)}E_{z}^{ \left( \omega \right)} \\ E_{x}^{ \left( \omega \right)}E_{z}^{ \left( \omega \right)} \\ E_{x}^{ \left( \omega \right)}E_{y}^{ \left( \omega \right)} \end{pmatrix}.
\end{equation}
Here, $\chi^{(2)}$ is the second-order nonlinear susceptibility. In the simulation of SHG, the undepleted pump approximation is adopted, assuming that the fundamental field remains nearly unaffected during the nonlinear conversion process. This procedure is typically divided into two steps. Firstly, the linear response of the structure is solved at the fundamental frequency to obtain the spatial distribution of the electric field at the nanoscale. Then, based on this field distribution, the nonlinear polarization source induced by the nonlinear effect is calculated and used as a driving source for the electromagnetic field calculation at the second harmonic frequency, thereby yielding the SHG radiation result. The SHG efficiency is defined as $\eta = P_{\text{SH}} / P_{\text{FF}}$, where $P_{\text{SH}}$ is the radiated power of the SH signal collected over a surface enclosing the entire scatterer, and $P_{\text{FF}}$ is the incident pump power at the fundamental wavelength. 

Fig. \hyperref[fig7]{7}(a) shows the wavelength-dependent behavior of SHG efficiency as a function of the internal nanocuboid parameter, under a maximum incident power of $0.78$ GW/cm$^{2}$. It can be observed that the SHG efficiency is significantly enhanced near the resonance wavelength, but rapidly diminishes as the excitation wavelength moves away from resonance. This behavior stems from the high-$Q$ factor and strong field localization enabled by the quasi-BICs mode. Fig. \hyperref[fig7]{7}(b) further compares the SHG efficiency versus wavelength for different values of the aspect ratio $H_2/D_2$. The results show that the system achieves optimal resonance at $H_2/D_2 = 1.54$, with a peak SHG efficiency reaching as high as $6.6\%$. As the parameter deviates from 1.54, the $Q$ factor of optical resonance reduces and the SHG efficiency correspondingly decreases. Additionally, an SHG enhancement peak can still be observed around 1650 nm, which is insensitive to the central nanocuboid's structural parameters, with an associated efficiency of approximately $0.3\%$. Fig. \hyperref[fig7]{7}(d) presents a bar chart of the maximum SHG efficiency for various values of $H_2/D_2$, showing excellent agreement with the structure’s linear optical response. We also investigated the dependence of SHG efficiency on pump power. Fig. \hyperref[fig7]{7}(e) illustrates the variation of SHG efficiency as the pump power increases from 0.1 GW/cm$^{2}$ to 1 GW/cm$^{2}$. The results indicate a linear relationship between SHG efficiency and pump power, confirming that the nonlinear conversion efficiency can be further improved by increasing the incident light intensity. Fig. \hyperref[fig7]{7}(c) compares the SHG efficiency enhancement of the triple nanocuboids with that of a single nanocuboid. The results demonstrate that the quasi-BICs resonance formed in the triple nanocuboids offers a significantly higher SHG enhancement than the single nanocuboid, and also far exceeds the MD resonance effect arising from the specific aspect ratio of the side nanocuboids in the coaxial triple nanocuboids configuration. Finally, we compare the SHG efficiency under AP and LP light excitations for the quasi-BICs mode. Fig. \hyperref[fig7]{7}(f) displays the wavelength-dependent SHG efficiency under AP and LP illumination at $H_2/D_2$ = 1.54. It is evident that the SHG efficiency under AP light is nearly four orders of magnitude higher than the counterpart under LP light, confirming the highly efficient coupling of AP light to the quasi-BICs mode, while LP light fails to effectively excite this mode.

\section{\label{sec:level5}Conclusions}

In conclusion, we demonstrate a high-$Q$ all-dielectric coaxial nanoresonators—composed of merely three subwavelength nanocuboids— that supports quasi-BICs capable of significantly enhancing local fields and nonlinear optical responses. Under AP illumination, the structure exhibits Fano resonances, and multipole analysis reveals that magnetic modes dominate the scattering process. In the nonlinear regime, the enhanced $Q$ factor facilitates substantial improvement in SHG efficiency, particularly under AP excitation, leading to a final enhancement of $6.6\%$ in SHG efficiency at a pump power of 0.76 GW/cm$^2$. The introduction of the mode overlap integral allows for a quantitative understanding of the coupling mechanism between incident fields and resonant modes. Far-field multipolar phase analysis further elucidates how destructive interference between radiation channels contributes to the suppression of radiative loss. This work not only demonstrates an effective strategy for achieving high-$Q$ resonances in compact all-dielectric systems but also provides theoretical tools and physical insight for guiding the design of advanced nonlinear photonic structures.

\begin{acknowledgments}
	
This work was supported by the National Natural Science Foundation of China (Grants No. 12304420, No. 12264028, No. 12364045, No. 12364049, No. 12474377, and No. 12104105), the Natural Science Foundation of Jiangxi Province (Grants No. 20232BAB201040, No. 20232BAB211025, and No. 20242BAB25041), the Young Elite Scientists Sponsorship Program by JXAST (Grants No. 2023QT11 and No. 2025QT04), and the Shanghai Pujiang Program (Grant No. 22PJ1402900).

\end{acknowledgments}

%\bibliography{Ref}


\end{document}